\documentstyle[12pt]{article}
  \advance\oddsidemargin  by -1in
  \advance\evensidemargin by -1in
  \advance\topmargin      by -0.5in
\newcommand {\bbox} {\vrule height7pt width4pt depth1pt}

\newcommand{\C}{{\cal C}}
\newcommand{\LL}{{\bf L}}
\newcommand{\HH}{{\bf H}}
\newcommand{\II}{{\rm I}}
\newcommand{\D}{{\bf D}}
\newcommand{\calR}{{\cal R}}

\newcommand{\calB}{{\cal B}}
\newcommand{\calF}{{\cal F}}
\newcommand{\calG}{{\cal G}}
\newcommand{\calL}{{\cal L}}
\newcommand{\calK}{{\cal K}}
\newcommand{\calM}{{\cal M}}
\newcommand{\calN}{{\cal N}}
\newcommand{\Ker}{\mathop{\rm Ker}\nolimits}
\newcommand{\Tr}{\mathop{\rm Tr}\nolimits}
\newcommand{\PR}{\mathop{\rm Pr}\nolimits}
\newcommand{\Prob}{\mathop{\rm Prob}\nolimits}
\newcommand{\trn}{\diamondsuit}
\newcommand{\ra}{\rangle}
\newcommand{\la}{\langle}
\title{Quantum Circuits with Mixed States}

\author{Dorit Aharonov\thanks{
Institutes of Physics and Computer science, The Hebrew University,
Jerusalem, Israel}
\and Alexei Kitaev\thanks{L.D.Landau Institute for Theoretical 
Physics,Moscow, Russia}
\and Noam Nisan\thanks{Institute of Computer science,
 The Hebrew University, Jerusalem,
Israel}}

\begin{document}
\maketitle
\thispagestyle{empty}

\begin{abstract}
\it{
Current formal models for quantum computation deal only with
unitary gates operating on ``pure quantum states''.  
In these models it is difficult or impossible to deal formally
with several central issues: measurements in the middle of the
computation; decoherence and noise, using probabilistic subroutines, and
more. It turns out, that the restriction to unitary gates and pure states is
 unnecessary. 
In this paper we generalize the formal model of quantum
circuits to a model in which the state can be a general quantum state, namely a 
mixed state,
or a ``density matrix'', and the gates can be general quantum operations, not 
necessarily 
unitary.
The new model is shown to be equivalent in computational power to 
the standard one, and the problems mentioned above essentially disappear.

The main result in this paper is a solution for the subroutine problem.
The general function that a quantum circuit outputs is a probabilistic 
function.  However, the question of using probabilistic functions as 
subroutines was not previously dealt with, the reason being that in
the language of pure states, this simply can not be done.
We define a natural notion of using general subroutines, 
and show that using general subroutines does not strengthen the model. 

As an example of the advantages of analyzing quantum complexity 
using density matrices, we prove a simple lower bound on depth of circuits
that compute  probabilistic functions.
Finally, we deal with the question of inaccurate quantum computation 
with mixed states. Using the so called ``trace metric'' on density matrices, 
we show how to 
keep track of errors in the new model. }

\end{abstract}

\vspace{1cm}
\newtheorem{theo}{Theorem}
\newtheorem{lemm}{lemma}
\newtheorem{conj}{conjecture}
\newtheorem{deff}{definition}
\newtheorem{coro}{corollary}

\section{Introduction}
In the last few years theoretical computer scientists have started
studying
``quantum computation''.  The idea, which originates with
Feynman \cite{Feyn82},
 may be 
summarized as
follows: the behavior of quantum physical systems 
seems to require exponential time to simulate on a (randomized) Turing
machine.  Hence, possibly, we could use physical computers built
by relying on quantum physical laws to get an exponential  speedup,
 for some computational problems, over normal (even randomized)
Turing machines. If true, then this contradicts what may be called
the modern 
Church-Turing thesis: ``randomized Turing machines
can simulate, with polynomial slowdown, any computation device''.

Deutsch formalized Feynman's idea, and defined computational models of
``quantum Turing 
machines''  and ``quantum circuits''
\cite{Deu}.  These are extensions of
the classical
models of Turing machines and circuits, that take into account the
laws of quantum physics.
 Deutsch's model, augmented with further work 
\cite{BV,Yao},
enabled a sequence of results 
\cite{BBBV,Sim,Gro}
 culminating with Shor's
polynomial quantum algorithm for factoring integers \cite{Shor94}.
However, it seems that Deutsch's
model is incomplete in some key aspects which make working formally
within it rather awkward.
It seems that there is still a gap between the physical world
 and the formal definitions,
which often leads computer scientists to bring 
physical phenomena into the model through the back door.

Let us recall the basic definitions used in current models
for quantum computation: The device operates on $n$ quantum-bits
(``qubits'').  Each one of the $2^n$
possible Boolean configurations $i \in \{0,1\}^n$ of these bits 
denotes a {\em basic state} $|i>$.  
The state of the computation at any point in time is a
{\em superposition} of basic states $\sum_{i \in \{0,1\}^n} c_i |i>$,
where the $c_i$'s are complex numbers satisfying $\sum_i |c_i|^2 =1$
(i.e. $\|c\|=\|c\|_2 =1$).
These superpositions are called {\em pure states}.  Each computational
operation is (1) {\em local:} i.e. involves only a constant number of 
qubits (2) {\em unitary:} i.e. maintains $\|c\|=1$.  At the end of the 
computation a {\em measurement} of one qubit is made, which returns 
a Boolean value ``true'' with probability $\sum_{i, |i> \in M} |c_i|^2$,
where $M$ is a subspace of $\cal{C}$$^{(\{0,1\}^n)}$ specified by
the measurement.   The state changes (in a well defined way) according
to the outcome of this measurement, and becomes a probability 
distribution over pure states.

Let us consider the model described above.
During the computation, the operations must be unitary, and the state must be a 
pure state.
But at the end of the computation, a non-unitary operation, a measurement, is 
applied,
and the state becomes a probabilisty distribution over pure states, or what is 
called a mixed
state. So we find out that in quantum physics operations might also be 
non-unitary,
and states are not nessecarily pure states. 
Restricting the model to unitary gates and pure states seems arbitrary.
It is natural to ask: what is the most general model that captures quantum 
physics? 
In this paper we define a quantum circuit which is allowed to be in 
a general quantum states,i.e. a mixed state, and which is allowed to use any 
 quantum operation as a gate, not necessarily unitary.
Our first result is a simple corollary of known results in physics: 
\cite{Sch}:

\begin{theo}\label{model}
 The model of quantum circuits with mixed states is polynomially equivalent in
computational power to the standard unitary model.
\end{theo}

There are several key issues, which we find  inconvenient, difficult, or 
impossible to deal with
inside the standard, unitary model. In the non-unitary model of quantum circuits 
with mixed states,
 these problems essencialy disappear. 
\begin{itemize}
\item
{\bf Measurements in the middle of computation}: it has been often remarked,
and implicitly used (e.g. Shor's algorithm), that quantum computations may allow 
measurements
 in the middle of the computation.  However, the state of the computation
after a measurement is a mixed state, which is not allowed in the standard 
model.
 This problem no longer exists in our 
model.

\item
{\bf Noise and Decoherence:} Noise and decoherence are key obstacles in 
implementing quantum computers devices.
Recent results show that theoretically,
 fault tolerant computation exists\cite{Sho96,AB97,KLZ,Kit97}.
Still, in the task of realizing quantum computers,
it is likely that more theoretical work will be needed in this direction.
A key problem in this  
  interface between quantum physics and 
quantum computation models is the fact that quantum noise,
and in particular, decoherence, are non-unitary operations that 
 cause a pure state to become a mixed state.
Incorporating quantum noise, which was impossible in the standard unitary 
model, 
is naturally done in our model.
\end{itemize}

The main technical result of this paper is a soulution of one more problem
 in the unitary model, namely the {\bf subroutine problem}.
 A cornerstone of computer science is the notion
of using subroutines (or oracles, or reductions): once we are able to
perform an operation $A$ within our model, we should be able to use $A$ as a 
``black box''  in further computations. 
The general and natural function that a quantum computer outputs is
a {\em probabilistic function:} for an input $X$, the output is distributed 
according to a 
distribution, $D_X$,
 which depends on the input. When using such a probabilistic function
 as a subroutine, the state is affected in a non-unitary manner. 
Therefore, using subroutines in their full generality was never defined in the 
unitary model.
 This is an incompleteness of the current model,
because computatble functions can not be used as ``black boxes''.

The special case of using  deterministic subroutines was defined
in\cite{BBBV}, and it was shown that using determinstic subroutines does not 
strengthen the model. As to probabilistic subroutines, these were
 implicitly used in quantum algorithms,( e.g. Shor's algorithm,)
but always on a classical input, for this case can be easily understood.
A conceptual difficulty lies in the combination of  
applying {\em probabilistic} subroutines to quantum {\em superpositions}.
It is not clear what the natural definition should be. Here,
 we are able to give  a natural definition  which 
generalises both the case
of deterministic subroutines on superpositions, and the case of
 probabilistic subroutines on classical inputs.
We prove that using general subroutines does not strengthen the model.
Let us define $FQP$ to be the set of probabilistic fuctions computed by uniform 
quantum circuits with
 polynomial size and depth. Our main result is:

\begin{theo}\label{subroutine} 
$FQP^{FQP}=FQP$.
\end{theo}

We hope that this new tool will be useful in quantum algorithms.

{~}

As to the formalism, it turns out that  
the description of the state of the circuit by a probability distribution over 
pure states 
is not unique. 
Physicists use an alternative unique description, namely density matrices,
which provides many conceptual and practical advantages.
In this paper, we give all definitions and proofs using the density matrix 
picture.
As an example of the benefits of dealing with 
quantum complexity questions with density matrices,
we provide a simple lower bound on the depth of a circuit 
which computes probabilistic functions.
The same lower bound seems difficult to prove when using the standard language 
of pure states.

A crucial point in a computation model is understanding how
inaccuracies in the basic computational elements
affect the correctness of the computed function.
In order to keep track of the error in the function
computed by a  circuit in a mixed state,
one needs appropriate metrics on probabilistic functions, density matrices
and gates.  
For probabilistic functions, we use a natural metric,relying on 
total variation distances.As a metric on density matrices, we 
propose to use the {\it trace metric,}
induced by the {\it trace norm}:
$\|H\|=\sum_i |\lambda_i|$,
where $ \lambda_i$ are the eigenvalues of $H$.
%
We show that it is an appropriate metric since 
 it quantifies the  {\em measurable distance between two quantum
states.} 
We also define a metric on quantum gates which has very nice properties.
An {\it error} in a function, a gate, or a density matrix will be 
the distance (in the corresponding metrics) from the correct function, gate, or 
density matrix, respectively. 
Using the above metrics, we  establish the following fact: 

\begin{theo}\label{errors}
Let $Q$ be a quantum circuit which uses  $L$ probabilistic subroutines
and gates, each  with at most 
 $\epsilon$ error.
The function that $Q$ computes has at most $O(L\epsilon)$ error.
\end{theo}

{\noindent}
{\bf Organization of Paper}
In section 2 we provide the physical background, in a mathematical language.
 In section 3 we define our model.  Section 4
provides the basic theorems regarding the model, and includes an example 
of complexity bound using density matrices. Section 5 
discusses the metrics and errors.

\section{Some Useful Physics Background}\label{phys}

The model of Quantum computers is based on the rules of quantum
mechanics. A good reference for basic rules is \cite{Saq}.

{\bf Pure states:}
A quantum physical system in a {\it pure state}
is described by a unit vector in a Hilbert space, 
i.e a vector space with an inner product.
In the {\em Dirac} notation a pure state is denoted by
$|\alpha\rangle$.
The physical system which corresponds to a quantum circuit
 consists of
 $n$ quantum two-state particles, and the
 Hilbert space of such a system is
 ${\cal H}_{2}^{n}= {\cal C}^{\{0,1\}^{n}}$
i.e. a $2^{n}$ dimensional complex vector space.
${\cal H}_{2}^{n}$ is viewed as a tensor product of $n$ Hilbert spaces
of one two-state particle: 
${\cal H}_{2}^{n}={\cal H}_{2}\otimes{\cal H}_{2}\otimes\ldots
\otimes{\cal H}_{2}$.
The $k$'th copy of ${\cal H}_{2}$ will be referred to as the  $k$'th
{\bf qubit}.
We choose a special basis for   ${\cal H}_{2}^{n}$,
which is called the computational basis.
It consists of the $2^{n}$ orthogonal states:
 $|i\rangle,0\le i< 2^{n}$, where $i$ is in binary representation.
$|i\rangle$ can be seen as a tensor product of states in
${\cal H}_{2}$: 
\(|i\rangle=|i_{1}\rangle|i_{2}\rangle....|i_{n}\rangle 
= |i_{1}i_{2}..i_{n}\rangle\),
where each $i_{j}$ gets 0 or 1.
Such a state, $|i\rangle$, corresponds to the j'th particle being in the
state    $|i_{j}\rangle$.
A pure state  $|\alpha\rangle\in{\cal H}_{2}^{n}$ 
is generally a {\em superposition}
 of the basis states:  
$|\alpha\rangle = \sum_{i=1}^{2^{n}} c_{i}|i\rangle$, 
 with $\sum_{i=1}^{2^{n}} |c_{i}|^{2}=1.$
A vector in  ${\cal H}_{2}^{n}$, 
 $v_{\alpha}=(c_{1},c_{2},...,c_{2^{n}})$, written in the computational
basis
representation, with \( \sum_{i=1}^{2^{n}} |c_{i}|^{2}=1\),  corresponds
to the pure state: 
\(|\alpha\rangle = \sum_{i=1}^{2^{n}} c_{i}|i\rangle\). 
 $v_{\alpha}^{\dagger}$, the transposed-complex conjugate of  $v_{\alpha}$,
is denoted  $\langle\alpha|$.
The inner product between $|\alpha\rangle$ and $|\beta\rangle$
is denoted $\langle\alpha|\beta\rangle=
 (v_{\alpha}^{\dagger},v_{\beta})$.
The matrix  $v_{\alpha}^{\dagger}v_{\beta}$
 is denoted as  $|\alpha\rangle\langle\beta|$.

{~}

{\bf Mixed state:}
In general, a quantum system is not in a pure state.
This may be attributed to the fact that
we have only partial knowledge about the system, 
or that the system is not isolated from the rest of 
the universe, so it does not have a well defined pure state.
We say that the system  
 is in a {\it mixed state}, and assign with the system
 a probability distribution, or {\em mixture} of 
pure states, denoted by $\{\alpha\}=\{p_{k},|\alpha_{k}\rangle\}$. 
This means that the system is with probability $p_{k}$ in the pure 
state $|\alpha_{k}\rangle$. 
 This description is not unique, as different
mixtures might represent the same physical system.
As an alternative description, physicists use the notion of 
{\bf density matrices}, which is an equivalent 
description but has many advantages.
 A density matrix $\rho$ on 
${\cal H}_{2}^{n}$ is an hermitian (i.e. $\rho=\rho^{\dagger}$)
semi positive definite matrix of dimension $2^n\otimes 2^n$ 
with trace $\Tr(\rho)=1$.
A pure state $|\alpha\rangle=\sum_{i} c_{i}|i\rangle$ 
is represented by the density matrix:
 \(\rho_{|\alpha\rangle} = |\alpha\rangle\langle\alpha|,\)\, i.e.  
\(\rho_{|\alpha\rangle}(i,j)= c_{i}c_{j}^{*}.\) (By definition, 
$\rho(i,j)=\langle i|\rho|j\rangle$).
A mixture $\{\alpha\}=\{p_{l},|\alpha_{l}\rangle\}$,
is associated with the density matrix\,
\(\rho_{\{\alpha\}} = \sum_{l} p_{l} |\alpha_l\rangle\langle\alpha_l|.\)
This association is not one-to-one, but it is {\bf onto} the
density matrices, because any density matrix describes
the mixture of it's eigenvectors, with the probabilities being 
the corresponding eigenvalues.
 Note that diagonal density matrices correspond to 
probability distributions over classical states.
 Density matrices are linear operators on their Hilbert spaces.
 The following notations will be used:
 if $\calN$ is a finite-dimensional
Hilbert space then $\LL(\calN)$ is the set of all linear operators on 
$\calN$. Also, $\LL(\calN,\calM)$
stands for the set of linear operators $\calN\to\calM$.

A density matrix of $n$ qubits can be reduced to a subset,
 $A$, of $m<n$ qubits. We say that 
the rest of the system, represented by the Hilbert space $\calF=\C^{2^{n-m}}$,
is {\em traced out}, and denote the new matrix by 
$\rho|_A=\Tr_{\calF}\rho$. It is defined as follows: 
\( \rho|_{A}(i,j)= \sum_{k=1}^{2^{n-m}} \rho(ik,jk)\). Actually, the 
partial trace $\Tr_{\calF}:\,\LL(\calN\otimes\calF)\to\LL(\calN)$
is defined for any pair of finite-dimensional Hilbert spaces $\calN$ and 
$\calF$. In words, it means averaging over $\calF$. Any quantum operation
which does not operate on $\calF$ commutes with this tracing out. 

{~}

{\bf Operations on quantum states}\label{op}
Transformations of density matrices are linear operators on operators
(sometimes called {\em super-operators}). Any physically allowed super-operator
$T:\,\LL(\calN)\to\LL(\calM)$ sends density matrices to density matrices. 
This is equivalent to say that $T$ is positive and trace-preserving. 
(A super-operator is called {\em positive} if it sends positive semi-definite
Hermitian matrices to positive semi-definite Hermitian matrices). However, 
this 
is not enough for a super-operator to be physically allowed. The positivity 
must remain if we extend the spaces $\calN$ and $\calM$ by adding more qubits.
That is, the super-operator $T\otimes\II_{\calF}$ must be positive, where 
$\II_{\calF}$ is the identity super-operator on an arbitrary finite-dimensional
Hilbert space $\calF$. Such $T$ is called a {\em completely positive map}.
Hence physically allowed quantum operations are linear
trace preserving completely positive maps.
Clearly, linear operations on mixed states 
preserve the probabilistic interpretation of the mixture, because
$T\circ\rho = T\circ
(\sum_{l} p_{l} |\alpha_l\rangle\langle\alpha_l|)=
\sum_{l} p_{l} T\circ( |\alpha_l\rangle\langle\alpha_l|).\)

One example for a super-operator 
 is the partial 
trace map which we defined before,
 $\Tr_{\calF}:\,\LL(\calN\otimes\calF)\to\LL(\calN)$. Another 
very important example is 
 a {\it unitary embedding} $V:\calN\to\calM$. This defines 
 the super-operator
$T: \rho\mapsto V\rho V^\dagger$. A unitary embedding 
naturally appears when we add a blank qubit to the system,\,
$V_0:\, |\xi\rangle\mapsto|\xi\rangle\otimes|0\rangle\,:
 \,\C^{2^n}\to\C^{2^{n+1}}$.\,
 It turns out that any
physically allowed super-operator is a combination of these two.

 The following lemma provides the link between super-operators and 
standard
unitary  operations.
It turns out that any super-operator from  $n$ to $m$ qubits 
is equivalent to the operation of a unitary matrix on 
$2n+m$ qubits.
\begin{lemm}\label{schuma}(modification of Choi (1970),Hellwig and Kraus(1975), and
Schumacher (1996)\cite{Sch}):
The following conditions are equivalent:
\begin{enumerate}
\item A super-operator $T:\LL(\calN)\to\LL(\calM)$ is trace-preserving 
and completely positive.
\item There is a Hilbert space $\calF$ with
 $\dim(\calF)\le\dim(\calN)\dim(\calM)$, and a unitary embedding 
$V:\,\calN\to\calN\otimes\calF$\, such that 
$T\rho=\Tr_{\calF}(V\rho V^{\dagger})
~\forall \rho\in \LL(\calN).$
\end{enumerate}
\end{lemm}

A super-operator corresponding to a unitary transformation on a space 
$\calN$,\,\, $|\alpha> \mapsto U|\alpha>$,\, sends a quantum state
$\rho=|\alpha\rangle\langle\alpha|$ to the state $U \rho U^{\dagger}$.
Such a super-operator is denoted by $U \cdot U^{\dagger}$. This is one
important of a physically realizable super-operator, 
and corresponds to the standard unitary operations. 

Super-operators can be extended to operate on larger spaces by
 taking tensor product with the identity operator:
$T:\LL(\calN)\to\LL(\calM)$ will be extended to 
$T\otimes I:\LL(\calN \otimes \calR)
\to\LL(\calM \otimes \calR)$.
Usually, we will be interested in those super-operators that are 
extensions of super-operators on spaces with small dimensionality.
This will correspond to local gates later on.

In order to describe the operation of 
 super-operators on density matrices $\in\LL(\calN) $  is suffices, from 
linearity, to
specify what happens to a basis which spans the  density matrices: 
$\{|i\rangle\langle  j|\}_{i,j}$ where $|i\ra, \la j|$ run 
over all basis vectors of $\calN$. (Any density matrix can be written as
  $\rho=\sum_{i,j} \rho_{i,j} 
|i\rangle\langle  j|$.)  
For example the unitary operation $|i\ra \longmapsto |v_i\ra$
corresponds to the super-operator specified by
$|i\ra\la j|\longmapsto |v_i\ra\la v_j|$.
If $U$ is extended to $\in\LL(\calN\otimes \calM)$
$|i,k\ra\la j,l|=|i\ra \la j|\otimes |k\ra \la l||\longmapsto   
(|v_i\ra\la v_j|)\otimes (|k\ra, \la l|).$

{~}

\noindent {\bf Measurements}
A quantum system can be {\bf measured}, or observed.
Let us consider a set of positive semi-definite Hermitian operators 
$\{P_m\}$, such that $\sum_m P_m=I$. The measurement is a process which 
yields a probabilistic classical output. For a given density matrix $\rho$,
the output is $m$ with probability $\PR(m)=\Tr(P_m\rho)$. 

We will use only {\em projection} measurements. Namely, we assume that $P_m$ 
are orthogonal projections onto mutually orthogonal subspaces $S_m$ 
which span the whole space $\calN=\C^{2^n}$, i.e.\ $\calN=\bigoplus_m S_m$. 
A more particular type of measurement which we 
will be using a lot is a basic measurement of $r$ qubits. In this case,
$P_m$ (with $1\le i\le 2^r$) are the projection on the subspace
$S_m$, which is the subspace spanned by basic vectors on which the measured 
qubits have the values corresponding to the string $m$:
$S_m={\rm span}\{\,|m,j\rangle,\  j=1,\ldots,2^{n-r}\}$.
This process corresponds to measuring the value of $r$ qubits,
in the basic basis, where here, for simplicity, we considered
measuring the first $r$ qubits.

The classical result of a measurement, $m$, can be represented by the density 
matrix $|m\rangle\langle m|$ in an appropriate Hilbert space $\calM$. The 
state of the quantum system after a projection measurement is also defined; 
it is equal to $\PR(m)^{-1}P_m\rho P_m$. (It has the same meaning as a 
conditional probability distribution). Thus, the projection measurement can be
described by a super-operator $T$ which maps quantum states on the space 
$\calN$ to quantum states on the space $\calN\otimes\calM$, the result being 
diagonal with respect to the second variable:
\[
  T\rho\ =\ \sum_m\, (P_m\rho P_m)\otimes\Bigl(|m\rangle\langle m|\Bigr)
\]

\section{Quantum Circuits with Mixed States}

We define here a model of quantum circuits, using density matrices.
This enables us to apply general non-unitary gates.
The circuit is defined to compute probabilistic functions, 
which are a generalization of Boolean functions computed by a standard 
quantum circuit. A quantum gate is defined to be the most general 
quantum operation:

\begin{deff}\label{defgate}
A {\bf quantum gate}, $g$, of order $(k,l)$ is a trace preserving, completely 
positive, linear map from density matrices on $k$ qubits to density matrices
on $l$ qubits. Its action on the density matrices is denoted as follows: 
\(\rho\longmapsto g\circ\rho\). 
(The ``$\circ$'' sign is used for clarity and could be omitted).
\end{deff}

The unitary gate, $U$, of the standard model is a special case of 
a quantum gate. The corresponding super-operator is $U\cdot U^{\dagger}$.
Using our ``$\circ$'' notation, we can denote it simply by $U$ with no danger
of confusion. Thus, \(U\circ\rho = U\rho U^{\dagger}\).

A measurement is also a special case of a quantum gate --- the probabilistic
projection onto a set of mutually orthogonal subspaces. Besides changing the
state of the qubits, it produces a classical probabilistic result.
(As shown above, both results can be described by a joint density matrix, 
but it is better to take 
the advantage of the second result being classical).

{~}

We now define a quantum circuit:

\begin{deff}
Let $\cal{G}$ be a family of quantum gates.
{\bf A Quantum circuit that uses gates from $\cal{G}$} 
is a directed acyclic graph. 
Each node $v$ in the graph is labeled by a gate $g_{v}\in\cal{G}$ of order
$(k_v,l_v)$. 
The in-degree and out-degree of $v$ are equal $k_v$ and $l_v$, respectively.
An arbitrary subset of the inputs are labeled  ``blank''.
An arbitrary subset of the outputs are labeled ``result''.
\end{deff}

 Here is a schematic example of such a circuit. 
We use ``b'', ``r'' for ``blank'', ``result''.

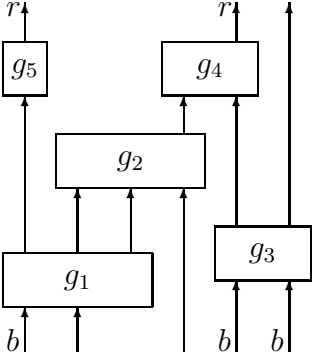
\begin{picture}(150,130)(-30,30)
\put(3,23){\shortstack{$b$}}
\put(10,23){\vector(0,1){17}}
\put(10,140){\vector(0,1){15}}
\put(3,150){\shortstack{$r$}}
\put(30,60){\vector(0,1){25}}
\put(30,23){\vector(0,1){17}}
\put(50,60){\vector(0,1){25}}
\put(2,40){\framebox(56,20){$g_{1}$}}
\put(10,60){\vector(0,1){60}}
\put(2,120){\framebox(16,20){$g_{5}$}}
\put(70,23){\vector(0,1){62}}
\put(70,105){\vector(0,1){15}}
\put(83,23){\shortstack{$b$}}
\put(90,23){\vector(0,1){27}}
\put(90,70){\vector(0,1){50}}
\put(90,140){\vector(0,1){15}}
\put(83,150){\shortstack{$r$}}
\put(110,23){\vector(0,1){27}}
\put(110,70){\vector(0,1){85}}
\put(103,23){\shortstack{$b$}}

\put(82,50){\framebox(36,20){$g_{3}$}}
\put(62,120){\framebox(36,20){$g_{4}$}}


\put(22,85){\framebox(56,20){$g_{2}$}}

\end{picture}

{~}

\noindent
The circuit operates on  a density matrix as follows:
\begin{deff} {\bf final density matrix}:
Let Q be a quantum circuit.
 Choose a topological sort
for Q: $ g_{t}...g_{1}$.
Where $g_{j}$ are the  gates used in the circuit.
The  final density matrix for an initial density matrix 
$\rho$ is $Q\circ \rho =
g_{t}\circ...\circ g_{2}\circ g_{1}\circ \rho$.
\end{deff}

Usually there exists more than one topological sort
for a given circuit Q.
Yet $Q\circ\rho$ is well defined  and operations of the gates in
 a quantum circuit determined by two different 
topological sorts are equivalent:
\begin{lemm}\label{topo}
$g_{t}\circ...\circ g_{2}\circ g_{1}\circ\rho
=g_{\sigma(t)}\circ...\circ g_{\sigma(2)}\circ g_{\sigma(1)}\circ\rho$,
where $\sigma$ is a permutation, 
If the two orderings are two topological sorts of the same quantum
circuit.
\end{lemm}

The reason for this is that two gates that operate on different qubits
commute. The proof of lemma \ref{topo} easily follows from:
\begin{lemm}\label{comu}
 Let $g_{1}, g_{2}$ be two quantum gates operating on different
qubits. Then  $\forall ~  \rho ~,~ g_{1}\circ g_{2}\circ\rho=g_{2}\circ
g_{1}\circ\rho$.
\end{lemm}

{\bf  Proof:} To extend the gates to operate on the whole 
set of $n$ qubits, we tensor with the identity. For simplicity, let us assume 
that $g_1$ operates on the first $k_1$ qubits (the Hilbert space $\calN$), 
and $g_2$ operates on some of the rest of the qubits 
(the Hilbert space $\calM$). We only need to show that\,
$(g_1\otimes\II_{\calM})\circ(\II_{\calN}\otimes g_2)\,=\,g_{1}\otimes g_{2}$
and\, 
$(\II_{\calN}\otimes g_2)\circ(g_1\otimes\II_{\calM})\,=\,g_{1}\otimes g_{2}$.
These are simply particular cases of the identity
\[
  (X_1\otimes X_2)\,(Y_1\otimes Y_2)\ =\ (X_1 Y_1) \otimes (X_2 Y_2)
\]
where $X_1,Y_1$ and $X_2,Y_2$ act on arbitrary linear two spaces. (The 
``$\circ$'' signs are omitted here).\bbox

Now we are ready to define the function that the circuit computes.
The circuit produces a probability distribution over strings
of $r$ bits, which depends on it's input string.
The probability distribution is computed out of the final density
matrix,
and is the same probability distribution as one would get
over the strings of outcomes, if at  the end of the computation
we apply a basic measurement of all the ``result''
qubits.

\begin{deff}\label{func}{\bf Computed function:}
Let Q be a quantum circuit, with $n$ inputs and $r$ ``result'' outputs. 
The probabilistic function that Q computes, {\bf $f_{Q}=f$}:
\( \{ 0,1 \}^{n} \longmapsto [0,1]^{\{0,1\}^{r}}, \) is defined as follows:
For input $i$, the probability for output $j$ is\, 
$f_{i,j}\, =\, 
 \langle j|\,\Bigl.\Bigl(Q\circ|i\rangle\langle i|\Bigr)\Bigr|_A\,|j\rangle$,\,
where $A$ is the set of the ``result'' outputs.
\end{deff}

\section{Results}

In this section we provide the theorems which prove that the model is 
equivalent
to the standard model and that it allows using probabilistic subroutines.
We also provide a simple lower bound on the depth of circuits 
computing probabilistic functions. 

\subsection{Substituting General Gates by Unitary Gates}

General non-unitary gates can be replaced by unitary gates,
as is shown by the following lemma:

\begin{lemm}\label{gate}
Let $g$ be a quantum gate of order $(n,m)$. There exists a unitary quantum 
gate $U_g$ on $2n+m$ qubits which satisfies:\, For any $\rho$,\,\,
$ g\circ\rho\,=\, \left.\left( U_g\circ 
  \Bigl(\rho\otimes |0^{n+m}\rangle\langle 0^{n+m}|\Bigr) \right)\right|_A$,\,
where $A$ is the set of the first $m$ qubits.
\end{lemm}

\noindent {\bf Proof:}
This lemma follows from lemma~\ref{schuma}.
By  definition~\ref{defgate}, the gate $g$ is a trace-preserving 
completely positive super-operator $\LL(\calN)\to\LL(\calM)$,\, where 
$\calN=\C^{2^n}$ and $\calM=\C^{2^m}$. By lemma~\ref{schuma}, it has 
a representation of the form $g=\Tr_{\calF}(V\cdot V^\dagger)$, 
where $\calF=\C^{2^{n+m}}$,\, and $V:\,\calN\to\calM\otimes\calF$ is a 
unitary embedding. Let us choose an orthonormal basis\,
$\{|\eta_{i,j}\rangle,\ 1\!\le i\!\le\!2^n,\ 1\!\le\!j\!\le\!2^{n+m}\}$\, in 
$\calM\otimes\calF$,\, 
such that $|\eta_{\,i,0^{n+m}}\rangle=V|i\rangle$ for any $i$\,
(the other basis vectors are arbitrary). There is a unique unitary operator 
$U$ which sends each vector $|i,j\rangle$ of the basic basis to the
vector $|\eta_{i,j}\rangle$ of the new basis. It is obvious that
$V=UV_0$, where $V_0:\, |\xi\rangle\mapsto|\xi\rangle\otimes|0^{n+m}\rangle$.
Hence $g=\Tr_{\calF}(UV_0\cdot V_0^\dagger U^{\dagger})$. This is what
we need. \bbox

{~}

\noindent
It follows that a circuit with general gates can be simulated by a circuit 
with unitary gates efficiently:

{~}

{\bf Theorem \ref{model}:} {\em 
 The model of quantum circuits with mixed states is
 polynomially equivalent in computational 
power to the standard model.}

\subsection{Using General Subroutines}
 A (probabilistic) subroutine is a function, $f: \{0,1\}^{m} \longmapsto 
R^{\{0,1\}^{p}}$,
which outputs $j$ with distribution which depends on the input $i$.
A quantum circuit that uses subroutines is a circuit
in which a node of fan-in=fan-out=$m+p$ may be associated instead of
a quantum gate, a probabilistic function $f: \{0,1\}^{m} \longmapsto 
R^{\{0,1\}^{p}}$.
Our definition of 
the way this ``subroutine gate'', denoted by $g_f$, effects the density matrix 
is 
by operating all possible deterministic functions, in the standard way,
where each deterministic function is applied with the induced probability from 
the probabilistic 
function:

\begin{deff}{\bf operation of a subroutine gate:}
\[g_f\circ\rho= \sum_d \PR(d) U_d\rho U_d^{\dagger}\]
Where the sum is over all deterministic functions $d: \{0,1\}^{m} \longmapsto 
\{0,1\}^{p}, ~U_d$ is defined by
$U_d|i,0>=|i,d(i)>$ and the induced probability for $d$ is:
           \[\PR(d)= 
\prod_i \Prob(i\mapsto d(i))=\prod_i f_{i,d(i)}.\]
\end{deff}

Note that as discussed in the introduction, 
 this definition generalizes 
both the case
of deterministic subroutines on superpositions, and the case of
 probabilistic subroutines on classical inputs.
The sum contains $2^{p^{2^m}}$ summands.
It turns out that the same operation on density matrices
 can be written in a much more compact form.

\begin{lemm}\label{gs}
\begin{eqnarray}
  g_f\circ(|i_1,0\rangle\langle i_2,0|)=~~~~~~~~~~~~~~~~~~~~~~~~~~~~~~~~~~~~~&\\ \nonumber   
\left\{ \begin{array}{ll}
    \sum_j f_{ij} |i,j\rangle\langle i,j|\quad & \mbox{if}\ i_1=i_2=i 
    \smallskip\\
    \sum_{j_1,j_2} f_{i_1j_1}f_{i_2j_2}\,|i_1,j_1\rangle\langle i_2,j_2|
    \quad & \mbox{if}\ i_1\not=i_2
    \end{array} \right. & 
 \nonumber \end{eqnarray}
\end{lemm}

\noindent{\bf Proof:}
\[\sum_d \PR(d) U_d|i_1,0\rangle\langle i_2,0| U_d^{\dagger}=\]
\[\sum_d \PR(d) |i_1,d(i_1)\rangle\langle i_2,d(i_2)|=\]\[
\sum_{j_1,j_2} \bigl(\sum_{d,d(i_1)=j_1, d(i_2)=j_2} \PR(d) \bigr) ~ 
|i_1,d(i_1)\rangle\langle i_2,d(i_2)|.\]

We now compute the term in the brackets.
If $i_1=i_2$,  it becomes:

\begin{eqnarray*}
\begin{array}{r}
\delta(j_1,j_2) \sum_{d,d(i_1)=j_1} \prod_i \Prob(i\mapsto d(i))=~~~~~~\\
\delta(j_1,j_2) f_{i_1,j_1}\sum_{j_i,i\not=i_1} \prod_i 
f_{i,j_i}=\delta(j_1,j_2)f_{i_1,j_1}.\end{array}\end{eqnarray*}
If $i_1\not=i_2$,
then the same computation yields
\begin{eqnarray*}
\begin{array}{r}
\sum_{d,d(i_1)=j_1,d(i_2)=j_2} \prod_i \Prob(i\mapsto d(i))=~~~~~~~~~\\
f_{i_1,j_1}f_{i_2,j_2}\sum_{j_i,i\not=i_1,i\not=i_2} \prod_i 
f_{i,j_i}=f_{i_1,j_1}f_{i_2,j_2}. ~~~\bbox \end{array}\end{eqnarray*}

Before we state our main theorem, here are some notations:
For a family of gates $\cal{G}$ we denote by $\cal{U_G}$ the set of unitary 
gates 
corresponding to gates from $\cal{G}$, according to lemma \ref{gate}.
We denote by $\cal{U^{\dagger}_{G}}$ the set of daggered unitary gates 
corresponding to gates from $\cal{G}$. $C$ is a special unitary gate on two 
qubits,
the  {\it controlled not gate}. It satisfies $C|00>=|00>,C|10>=|11>$,
and thus serves as a copying gate.

{~}

{\bf Theorem \ref{subroutine}:}
{\bf $FQP^{FQP}=FQP$}:
{\it  Let  $\cal{G}$ be a set of gates, $\cal{S}$ a set of 
probabilistic functions from $m$ to $r$ bits,  
 computable by  quantum circuits using no more than  $k$ gates from
  $\cal{G}$.
Let $Q$ be a quantum circuit, which uses $n$ gates from   $\cal{G}$,
 and $l$ subroutines from $\cal{S}$,}
{\it 
there exists a quantum circuit $\tilde{Q}$ which uses }
{\it no more then}
{\it $n+l(O(k)+O(r)+O(m))$ gates from }
 $\cal{U_G}$$\cup \cal{U_G^{\dagger}}\cup {C}$ {\it  and computes $f_{Q}$.}

{~}

{\bf Proof:} 
We now show that a subroutine $s\in S$, for  
$s: \{0,1\}^{m} \longmapsto R^{\{0,1\}^{r}}$,  can be replaced by  
$O(k)+O(r)+O(m)$ gates from  
 $\cal{U_G}\cup \cal{U_G^{\dagger}}\cup {C}$.
The idea is to apply $Q_s$, read the result by copying it to extra $r$ qubits,
 and undo the
subroutine. Up till now, this is just following the line of the 
 proof for deterministic subroutines\cite{BBBV}.  
However, this is not enough when dealing with probabilistic functions.
The reason, intuitively, is that in probabilistic subroutines there is more than 
one possible output
for one input, so the state of the bits that are used to copy the output,
 is not in tensor product with
that of the input and output bits, even if the input was classical. Hence, 
undoing the
subroutine does not take the input bits back to their original input state.
The reader is urged to try and see for herself why  more effort is needed.
We proceed by the following operations: We add $m+1$ blank qubits to the 
circuit.
The last bit will be a {\it garbage controll} bit.
First,  we check if there is garbage left, i.e. if the string written on the
qubits other than the main qubits is different from zero, and if so, 
we change the {\it garbage control} bit to $|1\rangle$.
Then, conditioned that the garbage control bit is one, we copy the {\em input} 
$m$ bit string 
of the subroutine to $m$ ancilla bits.
If there is no garbage,  we leave the ancilla bits to be blank.
Then we trace out, or discard, the garbage  and the $m+1$
ancilla bits. This procedure results with the same
operation as the subroutine gate, $g_s$, and we have used $O(k)+O(r)+O(m)$ 
gates.

Let us agree on some notation before continuing:
Let the subroutine $s$ be computed by the circuit $Q_s$, which uses only unitary 
gates
from  $\cal{U_G}$, using theorem \ref{gate}.
 Thus the operation of $Q_s$ is unitary, 
and is described by the unitary matrix  $U_s=U$. 
So the final density matrix of  $Q_s$ is a density matrix of a pure state,
which can be written as $U|i,0>$ for an input $i$, and $r$ blank qubits.
We can write:
\[  U|i,x\rangle\ =\ |i\rangle\otimes U_i|x\rangle  \]
\[  U_i|0\rangle\ =\ \sum_j\, |j\rangle\otimes|\psi_{ij}\rangle
    \qquad\quad \mbox{where}\quad
\langle\psi_{ij}|\psi_{ij}\rangle=f_{ij}
\]

Let us track the procedure step by step.
We will do that by seeing what happens to a matrix of the form
$|i_1\rangle\langle i_2|$. From linearity, this will be enough.

\[
    U|i,0\rangle\ =\ \sum_j\, |i\rangle\otimes|j,\psi_{ij}\rangle
\]

When $j$ is copied, the above expression becomes
\[ \sum_j\, |i,j\rangle\otimes|j,\psi_{ij}\rangle\]

The first two registers will be referred to as the {\em qubits},
and the last two registers will be discarded later.
Then $U^{-1}=U^\dagger$ is applied which yields

\[
    |\xi_i\rangle\ =\ 
    \sum_j\, |i,j\rangle\otimes\,U_i^\dagger|j,\psi_{ij}\rangle
\]

Let us represent $|\xi_i\rangle$ as $|\eta_i\rangle+|\nu_i\rangle$,
where
\[
    |\eta_i\rangle\ =\ 
    \sum_j\, \langle 0|U_i^\dagger|j,\psi_{ij}\rangle\, |i,j\rangle\otimes|0>
\]

corresponds to the possibility of having no garbage, and
$|\nu_i\rangle=|\xi_i\rangle-|\eta_i\rangle$ is orthogonal
to $|\eta_i\rangle$. Note that 
$\langle 0|U_i^\dagger|j,\psi_{ij}\rangle\,=\,
 \langle\psi_{ij}|\psi_{ij}\rangle\,=\,f_{ij}$,\, so
\[  |\eta_i\rangle\ =\ f_{ij}\, |i,j\rangle\otimes|0>\]
We now add the step of computing the state of the control garbage qubit, and 
conditioned on that coping the input.
The overall procedure can be represented as follows
\[ |i\rangle\ \mapsto\ |\eta_i,no,0\rangle + |\nu_i,yes,i\rangle\]
where $no$ and $yes$ are states of the controlled garbage bit.
As long as the garbage and the ancilla bits are discarded, i.e. the reduced 
density matrix 
on the original set of qubits (denote it by $Q$) is taken, we have:  
\[    |i_1\rangle\langle i_2|\ \mapsto\]
\[  (|\eta_{i_1},no,0\rangle + 
|\nu_{i_1},yes,i\rangle)
( \langle\eta_{i_2},no,0| + \langle\nu_{i_2},yes,i|)|_{Q}=\]\[=
    (|\eta_{i_1}\rangle\langle\eta_{i_2}|)|_{Q}+ 
(|\nu_{i_1}\rangle\langle\nu_{i_2}|)|_Q\delta_{i,j}\]

 For $i_1\not=i_2$, we have:

\[
 =\ 
    \sum_{j_1,j_2} f_{i_1j_1}f_{i_2j_2}\,|i_1,j_1\rangle\langle i_2,j_2|
\]

For $i_1=i_2$, we have to go few steps back in our calculations. Recall that 
the vectors $|\eta_{i}\rangle$ and $|\nu_{i}\rangle$ were defined in such a
way that $(|\eta_{i}\rangle\langle\nu_{i}|)|_Q=0$\, 
(because $|\eta_{i}\rangle$ corresponds to no garbage whereas $|\nu_{i}\rangle$
corresponds to non-null garbage). Hence
 \[(|\eta_{i}\rangle\langle\eta_{i}| + |\nu_{i}\rangle\langle\nu_{i}|)|_Q=
(|\xi_i\rangle\langle\xi_i|)|_Q\]
\[=\sum_{j,j'}\langle j',\psi_{ij'}|U_i U_i^\dagger|j,\psi_{ij}\rangle 
|i,j\rangle\langle i,j'|=\]
\[=\sum_{j} f_{ij} |i,j\rangle\langle i,j|.
\]


Thus we have the desired transformation. \bbox


\subsection{Simple Lower Bounds on Probabilistic Functions}
We prove a lower bound on probabilistic functions.
 The proof relies on {\it causality}, which can be stated as follows.
 Consider a quantum circuit $Q$, and two qubits $a$ 
and $b$. The two bits are correlated only if there is a gate 
from which there is a path to them both.
This will imply a lower bound on probabilistic functions where one qubit
is correlated to many others.

\begin{lemm}{\bf Causality lemma:}\label{causality}
Let $Q$ be a quantum circuit, with gates $g_t,...g_1$.
Let $\rho$ be a density matrix of a basic state. 
If $Q\circ\rho|_{a,b}$ is not a tensor product, 
there exist $i$ such that there are two  
(directed) paths in the circuit:  
$g_i \longmapsto a_f$ and $g_i \longmapsto b_f$.
\end{lemm}

{\bf Proof:}
Let us assume that there is no  $i$ such that there are two  
(directed) paths in the circuit:  
$g_i \longmapsto a_f$ and $g_i \longmapsto b_f$.
Let us now find a topological sort of all gates from which there is 
a directed path  
to $a_f$ (and therefore not to $b_f$), and let us call this set of gates $G_a$.
Let us sort the set $G_b$ similarly, and the rest of the gates $G_c$ also.
We claim that the sort $G_c G_b G_a$ is a topological sort of the circuit.
To show this, we need only show that if there is a path from $g_i$ to $g_j$ 
in the circuit, then $g_j$ appears to the left of $g_i$.
The only thing we have to check is that there is no path from gate $g_c$ in 
$G_c$
to any gate in $G_a$ ($G_b$).
But if there was such a path, then $g_c$ would have belonged to  $G_a$ ($G_b$).
Now,$(G_c\circ G_b\circ G_a\circ \rho)|_{a,b}=(G_b \circ G_a \circ\rho)|_{a,b}$
due to the following lemma:

\begin{lemm}\label{commute}
Let $g$ be a gate operating on qubits not in the set $B$.   
\(\rho|_{B}=(g\circ\rho)|_{B}. \)
\end{lemm}

{\bf proof:}
Let $B$ be described by first indices, 
and $g$ operates on the space described by second indices.
 \[\rho=\sum_{i,k,j,l} \rho_{ik,jl} |i,k\rangle\langle j,l|,~~~
\rho|_{B}=\sum_{i,j} (\sum_{k} \rho(ik,jk))|i\rangle\langle j|
\]
To apply the gate $g$, we use the equivalent unitary gate $U_g$ 
according to lemma \ref{gate}. 
\[g\circ\rho=\sum_{i,k,j,l}\rho_{ik,jl} |i\rangle\langle j| \otimes  
U_g|k\rangle\langle l|U_g^{\dagger}=\]
\[ \sum_{i,j,k,l,k',l'}\rho_{ik,jl} |i\rangle\langle j|\otimes 
|k'\rangle\langle k'|U_g|k\rangle \langle l|U_g^{\dagger} 
|l'\rangle\langle l'|.\]
Computing the reduced density matrix we get:
\[g\circ\rho|_{B}=\sum_{i,j}|i\rangle\langle 
j|\bigl(\sum_{k,l,k'}\rho_{ik,jl}\langle k'
| U_g|k\rangle \langle l|U_g^{\dagger} |k'\rangle,\bigr)\]
but 
\(\sum_{k'} \langle k'| U_g|k\rangle\langle l|U_g^{\dagger} |k'\rangle=\langle 
l|U_g U_g^{\dagger}|k\rangle=\delta_{k,l}. \bbox\)


The set of qubits $A$, $B$ which $G_a$ and $G_b$ operate upon
are disjoint, according to our assumption.
Let $A',B'$, be sets of qubits such that their union is all the qubits,
and $A'\supseteq A$,
$B'\supseteq B$. We can write:
\[
(G_b \circ G_a \circ\rho)|_{a,b}=
(G_b \circ G_a \circ(\rho|_{A'}\otimes \rho|_{B'}))|_{a,b}=\]
\[(G_a\circ\rho|_{A'})\otimes(G_b \circ\rho|_{B'})|_{a,b}=
(G_a\circ\rho|_{A'})|_a \otimes(G_b \circ\rho|_{B'})|_{b}.\]

Which shows that the final reduced density matrix is a tensor product. \bbox

Let us define the correlation graph for a state:
\begin{deff}{\bf Correlation graph:}
Given a state $\rho$ of $n$ qubits, we define the correlation graph 
$G_{\rho}(V,E)$ of the state as
follows.
The set of nodes $V$ will consist of  $n$ nodes,
corresponding to the $n$ qubits.
An edge $(a,b)\in E$ iff the reduced density matrix $\rho|_{a,b}$
is not a tensor product.
\end{deff}
We now claim that the depth of the circuit with final density matrix $\rho$
is larger than the logarithm of the maximal degree in the correlation graph 
$G_{\rho}$.
\begin{lemm} 
Let $Q$ be a quantum circuit, with all gates of fan-in $\le k$.
Let the maximal degree of the correlation graph of $Q\circ |i>$ be $c$,
for some input $i$. 
Then the depth of $Q$ satisfies $ D(Q)\ge \frac{1}{2}log_k(c)$.
\end{lemm}

{\bf Proof:}
By causality, if there is an edge in the correlation graph between qubits
$a,b$ then in the circuit there is a node 
 $g_i$ such that there are two  
(directed) paths in the circuit:  
$g_i \longmapsto a_f$ and $g_i \longmapsto b_f$.
For a circuit of depth $D$, and a given qubit $a$,
the maximal number of qubits which are connected to $a$ in such a way are 
$k^{2D}$.
So $c\le k^{2D}$, and hence $D(Q)\ge \frac{1}{2} log_{k}(c)$.\bbox

The correlation graph can be defined for probabilistic functions as well.
If the output is probabilistic string of $r$ bits, it will be a graph of $r$ 
nodes.
Edges will connect pairwise correlated bits.  
\begin{lemm} {\bf Correlation bound:}
Let $Q$ be a quantum circuit computing $f$, a probabilistic function.
Let $c$ be the maximal degree of the correlation graph of $f$.
Then $D(Q)\ge log_k(c)$.
\end{lemm}

As a trivial example, consider the probabilistic function
that outputs (for any input) with probability $\frac{1}{2}$  the string $0^r$ 
and with probability   $\frac{1}{2}$  the string $1^r$.
The lemma shows that a circuit that computes this function must be of 
depth larger than $log(r)$.

\section{Precision and Errors}

In the theory of quantum computation (as in the real life) operators, quantum
states, etc.\ are defined with some precision. Thus, we need to define certain
metrics on the corresponding spaces. We will find a natural metric (more
specifically, a norm) for each class of objects we deal with: pure and mixed
states, unitary and arbitrary gates. After proving some basic properties of
these norms, we will show, in a very general form, that error accumulation
in quantum computation is at most additive (see Theorem~\ref{th_error} below).

\subsection{The Natural Distance Between Probabilistic Functions}
\noindent We need a measure for the accuracy of the function computed. 
The natural norm to use is the  $\ell_1$-norm,\, 
called the total variation distance (t.v.d.)
 between probability distributions.
We use t.v.d. to  define a metric on probabilistic functions. 
\begin{deff}
Let $f,g$ be probabilistic functions. For input $i$,
 $f_i,g_i$ are probability distributions. 
The total variation distance between $f_i,g_i$ is
\(
|f_i-g_i|\stackrel{\rm def}{=} \sum_{j} |f_{i,j}-g_{i,j}|\) and
 \(\|f-g\| = max_{i}|f_i-g_i|.\) 
\end{deff}



\subsection{The Trace Metric on Density Matrices}

Precision of a vector $|\xi\rangle\in\calN$ (where $\calN$ is a Hilbert space)
is characterized by the natural (Euclidean) norm 
$\|\xi\rangle\|=\sqrt{\langle\xi|\xi\rangle}$.\,
Since we have passed to density matrices, we need a metric
on general quantum states.
There are two natural norms on the space of linear operators on $\calN$:\, 
the usual operator norm,
\begin{equation}
  \|A\|\ =\ \sup_{|\xi\rangle\not=0} \frac{\|A|\xi\rangle\|}{\|\xi\rangle\|}
  \,\ =\,\ \mbox{largest eigenvalue of}\ \sqrt{A^\dagger A}
\end{equation}
\noindent and the dual norm called the {\em trace norm},
\begin{equation}
  \|A\|_{1}\ =\
  \sup_{B\not=0}  \frac{|\Tr AB|}{\|B\|} \ =\
  \Tr\sqrt{A^\dagger A}\,\ ~~~~~~~~~~~~~~~~
\end{equation}

The norms $\|\cdot\|$ and $\|\cdot\|_1$ are well behaved.
 Specifically, if $A\in\LL(\calN)$ and $B\in\LL(\calM)$ then

\begin{lemm}\label{normprop}
\begin{equation}\begin{array}{l}
  \|A\otimes B\|\, =\, \|A\|\,\|B\|, \\
  \|A\otimes B\|_1\, =\, \|A\|_1\,\|B\|_1\\
  \|AB\|\, \le\, \|A\|\,\|B\|\\
  \|AB\|_{1},\, \|BA\|_{1}\ \le\ \|B\|\,\|A\|_{1} \\
  |\Tr A|\, \le\, \|A\|_{1}\end{array}
\end{equation}
\end{lemm}
(Proof is trivial).

{~}
There are many good reasons to use the trace norm as the norm 
on density matrices (though the operator norm will be very 
useful in proofs.)
First, two pure states $|\xi\rangle,|\eta\rangle$
which are close in the Euclidean norm are close also in the trace norm:
\[
  \Bigl\| |\xi\rangle\langle\xi| - |\eta\rangle\langle\eta| \Bigl\|_{1}\ =\
  2\,\sqrt{1-|\langle\xi|\eta\rangle|^2}\ \le\
  2\,\Bigl\| |\xi\rangle - |\eta\rangle \Bigr\|
\]
The important feature of the trace metric is that it  
 captures the {\it measurable} distance 
between different density matrices.
It turns out that the trace distance between two density matrices 
equals the following quantity. For each observable $O$,
a density matrix $\rho$ induces a probability
distribution,  $p^{O}_{\rho}$, over $i's$.
The trace distance between two density matrices is the maximal t.v.d
between the two probability distributions, taken 
over all possible observables.
\begin{lemm}\label{QmetP}
\(\|\rho_{1}-\rho_{2}\|_1=
 max_{O} \left\{|p^{O}_{\rho_{1}}-p^{O}_{\rho_{2}}|\right\}.\)
\end{lemm}
\noindent{\bf Proof:}  
Let $\calN=\bigoplus_{j}S_j$, where the subspaces $S_j$ are
mutually orthogonal. Let $P_j$ be the orthogonal projection onto $S_j$.
Then, for any pair of mixed states $\rho_1$ and $\rho_2$,
\(
  \sum_{j} \Bigl| \Tr(P_j\rho_1) - \Tr(P_j\rho_2) \Bigr|\ \le\
  \|\rho_1-\rho_2\|_{1}
\). To see this, present the left hand side
 of this inequality  as
$\Tr((\rho_1-\rho_2)B)$,\, where\, $B=\sum_{j}\pm P_j$.\,
It is obvious that $\|B\|= 1$. Then use lemma~\ref{normprop}. 
To see that the trace distance can be achieved by some measurement, 
let $O$ project on  the eigenvectors of  $\rho_1-\rho_2$.
\bbox

\subsection{The Diamond Metric on Quantum Gates}

\noindent The natural norm on the space of super-operators is
\[
  \|T\|_1\ =\ \sup_{X\not=0}\frac{\|TX\|_1}{\|X\|_1}
\]
Unfortunately, this norm is not stable with respect to tensoring with the 
identity. Counterexample:\, 
$T:\,|i\rangle\langle j|\mapsto|j\rangle\langle i|$\,\, ($i,j=0,1$).\, 
It is clear that $\|T\|_1\le 1$. However $\|T\otimes I_\calB\|_1\ge 2$. 
(Apply the super-operator $T\otimes I_\calB$ to the operator 
$X=\sum_{i,j}|i,i\rangle\langle j,j|$~).
For this reason, we have to define another norm on super-operators
\begin{deff}\label{trnorm} Let $T:\LL(\calN)\to \LL(\calM)$
and  
$A,B\in\LL(\calN,\,\calM\otimes\calF)$, where $\calF$ is
 an arbitrary Hilbert space of dimensionality
$\ge(\dim\calN)(\dim\calM).$
\[
  \|T\|_\trn\ =\
  \inf\Bigl\{ \|A\|\,\|B\|\,:\ \Tr_{\calF}(A\cdot B^\dagger)=T \Bigr\}
\]
\end{deff}
This definition seems very complicated. However it is worthwhile 
using this norm because it satisfies very nice properties, and provides 
powerful tools for proofs regarding quantum errors. 
Here are some properties which are satisfied by the diamond
  norm. The first property is
that the diamond norm is the stabilized version of the ``naive'' 
norm $\|\cdot\|_1$. The proof of this is complicated and non-trivial.
It implies also that $\|\cdot\|_\trn$ is a norm.

\begin{lemm}\label{trnormprop}
\end{lemm}
\begin{enumerate}
\item\label{stabnorm} \(
 \|T\|_\trn\,=\,\|T\otimes I_\calG\|_1\,\ge\,\|T\|_1,~
where, \dim\calG\ge\dim\calN.\)
\item\label{oper} \( \|T\rho\|_\trn\ \le\ \|T\|_\trn\,\|\rho\|_1 \)
\item\label{prod} \(  \|TR\|_\trn\ \le\ \|T\|_\trn\,\|R\|_\trn \)
\item\label{tens} \( \|T\otimes R\|_\trn\ =\ \|T\|_\trn\,\|R\|_\trn\)
\item\label{unit}
 The norm of any physically allowed super-operator $T$ is equal to $1$. 
\item If $\|V\|\le 1$ and $\|W\|\le 1$ then\,
$\|V\cdot V^\dagger-W\cdot W^\dagger\|_\trn\le 2\|V-W\|$.
\end{enumerate}

\noindent{\bf Proof of \ref{trnormprop}.1}:
It is easy to see that\, $\|T\otimes I_\calG\|_1\le\|T\otimes
I_\calG\|_\trn\le\|T\|_\trn$.\, 
The inequality $\|T\|_\trn\le\|T\otimes I_\calG\|_1$ is not so obvious.  
W.l.o.g.\ $\|T\|_\trn=1$. We are to prove that $|T\otimes I_\calG\|_1\ge1$.

We will use the following notation. $\D(\calK)$ denotes the set 
of density matrices on $\calK$, whereas $\HH(\calF)$ is 
the set of Hermitian operators.
We can impose the restriction $\|A\|=\|B\|\le 2$ without changing the 
infimum in the definition~\ref{trnorm}. Due to compactness, the infimum is 
achieved at some $A$ and $B$. W.l.o.g.\ $\|A\|=\|B\|=1$. 
The quantity $\|A\|\,\|B\|$ is minimal with respect to infinitesimal 
variations of the scalar product 
$\delta\langle\cdot|\cdot\rangle=\langle\cdot|Z|\cdot\rangle$ on the space
$\calF$. (Here $Z$ is a infinitely small Hermitian operator on $\calF$).
When computing the variations $\delta\|A\|$ and $\delta\|B\|$,\, 
we can restrict $A$ and $B$ to the subspaces $\calK=\Ker(A^\dagger A-1_\calN)$ 
and $\calL=\Ker(B^\dagger B-1_\calN)$. Clearly,
\[
  \begin{array}{rclcl}
  \delta\|A\| &=& \max_{|\xi\rangle\in\calK,\,\,\|\xi\rangle\|=1}\limits\,
  \langle\xi|A^\dagger(1_\calM\otimes Z)A|\xi\rangle   \\&&\\
 &=& \max_{X\in E}\limits \Tr(XZ)
  \bigskip\\
  \delta\|B\| &=& \max_{|\eta\rangle\in\calL,\,\,\|\xi\rangle\|=1}\limits\,
  -\langle\eta|B^\dagger(1_\calM\otimes Z)B|\eta\rangle\\&&\\
  &=& \max_{Y\in F}\limits\, -\Tr(YZ)
  \end{array}
\]
where
\[
  E\ =\ \Bigl\{ \Tr_\calM(A\rho A^\dagger):\,\rho\in\D(\calK) \Bigr\}\]
\[  F\ =\ \Bigl\{ \Tr_\calM(B\gamma B^\dagger):\,\gamma\in\D(\calL) \Bigr\}
\]
Thus, for any $Z\in\HH(\calF)$\,
\[
  \delta\,\Bigl(\|A\|\,\|B\|\Bigr)\ =\ 
  \max_{X\in E,\,\,Y\in F}\,(\Tr(XZ)-\Tr(YZ))\ \ge\ 0
\]
This means that the sets $E,F\subseteq\HH(\calF)$ can not be separated by a 
hyper-plane. As $E$ and $F$ are convex and compact,\, $E\cap F\not=\emptyset$.
Let $\Tr_\calM(A\rho A^\dagger)=\Tr_\calM(B\gamma B^\dagger)\in E\cap F$,\, 
where\, $\rho\in\D(\calK)$,\,\, $\gamma\in\D(\calL)$.  Let us represent 
$\rho$ and $\gamma$ in the form\, 
$\rho=\Tr_\calG\Bigl(|\xi\rangle\langle\xi|\Bigr)$,\,\,
$\gamma=\Tr_\calG\Bigl(|\eta\rangle\langle\eta|\Bigr)$,\, where\,
$|\xi\rangle,|\eta\rangle\in\calN\otimes\calG$ are unit vectors.
Put $X=|\xi\rangle\langle\eta|$. Then\, $\|(T\otimes I_\calG)X\|_1=\|X\|_1=1$

{~}

\noindent {\bf Proof of \ref{trnormprop}.2,\ref{trnormprop}.3:}
 follow from the relation to the norm $\|\cdot\|_1$
and the definition of the latter.

{~}

\noindent {\bf Proof of \ref{trnormprop}.4}: To prove first direction, 
$\|T\otimes R\|_\trn\ \le\ \|T\|_\trn\,\|R\|_\trn$ follows from the 
definition~\ref{trnorm}, whereas the inverse inequality follows from
\ref{stabnorm}.

{~}

\noindent{\bf Proof of \ref{trnormprop}.5}:
W.~l.~o.~g.\ $\|T\|_\trn=\|T\|_1$ (since we can tensor $T$ with the identity).
Let $T=\Tr_{\calF}(V\cdot V^\dagger)\,:\,\LL(\calN)\to\LL(\calM)$. Let us
define the dual super-operator $R:\LL(\calM)\to\LL(\calN)$ with the property:
$\Tr(Y(TX))=\Tr((RY)X)$ for every $X\in\LL(\calN)$ and $Y\in\LL(\calM)$. 
It is obvious that $RY=V^\dagger(Y\otimes I_{\calF})V$ and
\[
  \|T\|_1\ =\ \sup_{Y\not=0}\frac{\|RY\|}{\|Y\|}
\]
As $\|V\|\le 1$, the inequality $\|T\|_1\le 1$ follows immediately. To prove
the inverse inequality, take the identity operator for $Y$.

{~}

\noindent{\bf Proof of \ref{trnormprop}.6}: 
To prove this we need lemma \ref{adderror} from the next section.
(For completeness this property appears here).
 In the definitions of the lemma put
 $T_1=V\cdot I$,\,\, $T_1'=W\cdot I$,\,\, 
$T_2=I\cdot V^{\dagger}$,\,\, $T_2'=I\cdot W^{\dagger}$.~~\bbox

{~}

The distance between unitary super-operators $V\cdot V^{\dagger}$,$W\cdot
W^{\dagger}$,   can be calculated explicitly, 
and has a geometrical interpretation.
Denote by\, $d$\, the distance between $0$ and the polygon 
(in the complex plane) whose vertices are the eigenvalues
of $VW^\dagger$. Then 
\[\|V\cdot V^\dagger-W\cdot W^\dagger\|_\trn =~~~~~~~~~~~~~~~~~~ \] 
\[  \max_{\rho\in\D(\calN)} \Bigl\| 
V\rho V^\dagger-W\rho W^\dagger \Bigr\|_1
 =\ 2 \sqrt{1-d^2}.
\]
(The proof is left to the reader).

\subsection{Bounding the Overall Error}

By definition, the error of a quantum gate is measured by the $\trn$-norm.
 The accumulation of errors is bounded by the following lemma:

\begin{lemm} \label{adderror}
Let $T_1$, $T_2$ and $T_1'$, $T_2'$ be super-operators with norm $\le 1$,\,
such that\, $\|T_j'-T_j\|_\trn\le\epsilon_j$\,\, ($j=1,2$). 
Then $\|T_2' T_1'-T_2 T_1\|_\trn\,\le\,\epsilon_1+\epsilon_2$.
\end{lemm}
\noindent{\bf Proof:} Write
$T_2' T_1'-T_2 T_1\,=\, T_2'(T_1'-T_1)+(T_2'-T_2)T_1$, and use lemma 
\ref{trnormprop}.5 \bbox

For subroutines, the natural error measure 
is different. Fortunately, there is a linear upper bound for the $\trn$-norm 
error of a subroutine:

\begin{lemm}\label{ersub}
Let $f$ and $f'$ be two 
probabilistic subroutines, such that $\|f-f'\|\le \epsilon$. 
Then $\|g_{f'}-g_{f}\|_\trn\le\,5\epsilon$. 
(The super-operator $g_{f}$ is described in the lemma~\ref{gs}).
\end{lemm}

\noindent{\bf Proof:}
Let $\calN$ be the space of the inputs $|i\rangle$,\, $\calM$ the space of
the outputs $|i,j\rangle$. Define the following objects (and their primed 
versions)
\[
\begin{array}{rclcl}
  A &:& \calN\to\calM &:& |i\rangle\,\mapsto\,\sum_{j}f_{ij}|i,j\rangle
  \medskip\\
  B &:& \calN\to\,\calM\otimes\calN &:& 
  |i\rangle\,\mapsto\,\sum_{j}f_{ij}|i,j,i\rangle
\end{array}
\]
\[
  \rho_i\ =  \sum_{j}f_{ij}|i,j\rangle\langle i,j|\]\[
  P_i\ :\ \LL(\calN)\to\C\ :\ |i_1\rangle\langle i_2|\mapsto\delta_{i_1i_2}
\]
Then $g_{f}=A\cdot A^\dagger-\Tr_{\calN}(B\cdot B^\dagger)+\sum_i\rho_i P_i$
(the same for $g_{f'}$). Clearly,\, $\|A'-A\|\le\epsilon$,\,\,
$\|B'-B\|\le\epsilon$,\, and the norm of each operator $A$, $B$, $A'$, $B'$
does not exceed $1$. It remains to show that $\|T\|_\trn\le\epsilon$, where
$T=\sum_i(\rho_i'-\rho_i)P_i$.

Note that $\|\rho_i'-\rho_i\|_1\le\epsilon$ for each $i$. Hence\, 
$\|(T\otimes I_{\calG})X\|_1\,\le\, \epsilon\,\sum_i\|Y_i\|_1$,\, 
where $\calG$ is an arbitrary Hilbert space,\, $X\in\LL(\calN\otimes\calG)$,\,
and $Y_i=(P_i\otimes I_{\calG})X$. On the other hand, 
\[
  \sum_i\|Y_i\|_1\ =\ 
  \left\| \sum_i |i\rangle\langle i|\otimes Y_i \right\|_1\ =\
  \|(P\otimes I_{\calG})X\|_1\ \]\[ \le\ \|P\|_\trn\,\|X\|_1\ =\ \|X\|_1
\]
where\, 
$P:\ |i_1\rangle\langle i_2|\mapsto\,\delta_{i_1i_2}|i_1\rangle\langle i_2|$
is a physically realizable super-operator.
\bbox

Due to lemmas \ref{adderror} and \ref{QmetP}, an $\epsilon$ error 
generated somewhere in the circuit can not contribute more than 
$\epsilon$ error to the computed function.
This proves the following theorem:

\begin{theo}\label{th_error}
Let $Q$ be a quantum circuit which uses  $L$ probabilistic subroutines
and gates, each  with at most $\epsilon$ error.
The function that $Q$ computes has at most $O(L\epsilon)$ error.
\end{theo}

\section{Acknowledgments}
We thank Michael Ben-or and Avi Wigderson for valuable remarks.
Part of this work was completed during the 1997 Elsag-Bailey -- I.S.I. 
 Foundation research meeting on quantum computation.
One of us (A. Kitaev) is supported 
by the Russian Foundation for Fundamental Research (grant 96-01-01113) and
the Landau-ENS cooperation program.

\small
\bibliographystyle{plain}
\bibliography{references}

\end{document}